\documentclass[11pt,twoside]{article}
\usepackage{asp2014}

\usepackage[usenames,dvipsnames]{xcolor}

\aspSuppressVolSlug
\resetcounters

\begin{document}

\title{Weak gravitational lensing with CO galaxies}
\author{Philip Bull
\affil{Radio Astronomy Laboratory, University of California Berkeley, Berkeley, CA 94720, USA; \email{philbull@berkeley.edu}} }

\author{Ian Harrison
\affil{Jodrell Bank Centre for Astrophysics, School of Physics \& Astronomy, The University of Manchester, Manchester M13 9PL, UK; \email{ian.harrison-2@manchester.ac.uk}} }

\author{Eric Huff
\affil{Jet Propulsion Laboratory, California Institute of Technology, 4800 Oak Grove Drive, Pasadena, California, USA; \email{Eric.M.Huff@jpl.nasa.gov }} }

\paperauthor{Philip Bull}{philbull@berkeley.edu}{0000-0001-5668-3101}{University of California Berkeley}{Department of Astronomy}{Berkeley}{CA}{94720}{USA}
\paperauthor{Ian Harrison}{ian.harrison-2@manchester.ac.uk}{}{Jodrell Bank Centre for Astrophysics}{University of Manchester}{Manchester}{}{M13 9PL}{UK}
\paperauthor{Eric Huff}{Eric.M.Huff@jpl.nasa.gov}{}{Jet Propulsion Laboratory}{California Institute of Technology}{Pasadena}{CA}{91109}{USA}

\begin{abstract}
Optical weak lensing surveys have become a powerful tool for precision cosmology, but remain subject to systematic effects that can severely bias cosmological parameter estimates if not carefully removed. We discuss the possibility of performing complementary weak lensing surveys at radio/microwave frequencies, using detections of CO-emitting galaxies with resolved continuum images from ngVLA. This method has completely different systematic uncertainties to optical weak lensing shear measurements (e.g. in terms of blending, PSF, and redshift uncertainties), and can provide additional information to help disentangle intrinsic alignments from the cosmological shear signal. A combined analysis of optical and CO galaxy lensing surveys would therefore provide an extremely stringent validation of highly-sensitive future surveys with Euclid, LSST, and WFIRST, definitively rejecting biases due to residual systematic effects. A lensing survey on ngVLA would also provide valuable spectral (kinematic) and polarimetric information, which can be used to develop novel cosmological analyses that are not currently possible in the optical.
\end{abstract}

\section{Introduction and scientific motivation}

As light travels from distant galaxies, it is distorted by the gravitational pull of matter that it encounters along the way. While some light rays pass close enough to dense concentrations of matter to be strongly lensed, producing arcs and multiple images, the majority of rays are only slightly affected. This weak gravitational lensing effect can be measured in a number of ways, but the most common is to look for lensing shear, which causes coherent correlations in the ellipticity of galaxies over angular scales of a few arcminutes or larger \citep{Bartelmann:1999yn, Munshi:2006fn, Hoekstra:2008db, Kilbinger:2014cea, 2017arXiv171003235M}. This has motivated the development of a slew of large galaxy surveys, predominantly in the optical, that seek to measure the shapes of many tens or even hundreds of millions of galaxies across cosmic time in order to characterize the statistical properties of the weak lensing `shear' signal.

Weak lensing has rapidly developed into one of the pillars of observational cosmology. This is due in large part to its extremely high information content -- the lensing signal not only encodes information about the total matter distribution (including cold dark matter and baryons) along the line of sight, but also the growth rate of large-scale structure and the geometry of space \citep{VanWaerbeke:1998tu, Jain:2003tba}. These quantities are instrumental in attempts to understand some of the biggest problems in cosmology and fundamental physics, such as the nature of cosmic acceleration and the Cosmological Constant problem, and the nature of gravity on extremely large distance scales. Lensing can also be used to understand the connection between galaxies and the dark matter structures that they are embedded in \citep{Hoekstra:2003pn, Sheldon:2003xj}.

\articlefigure[width=0.7\textwidth]{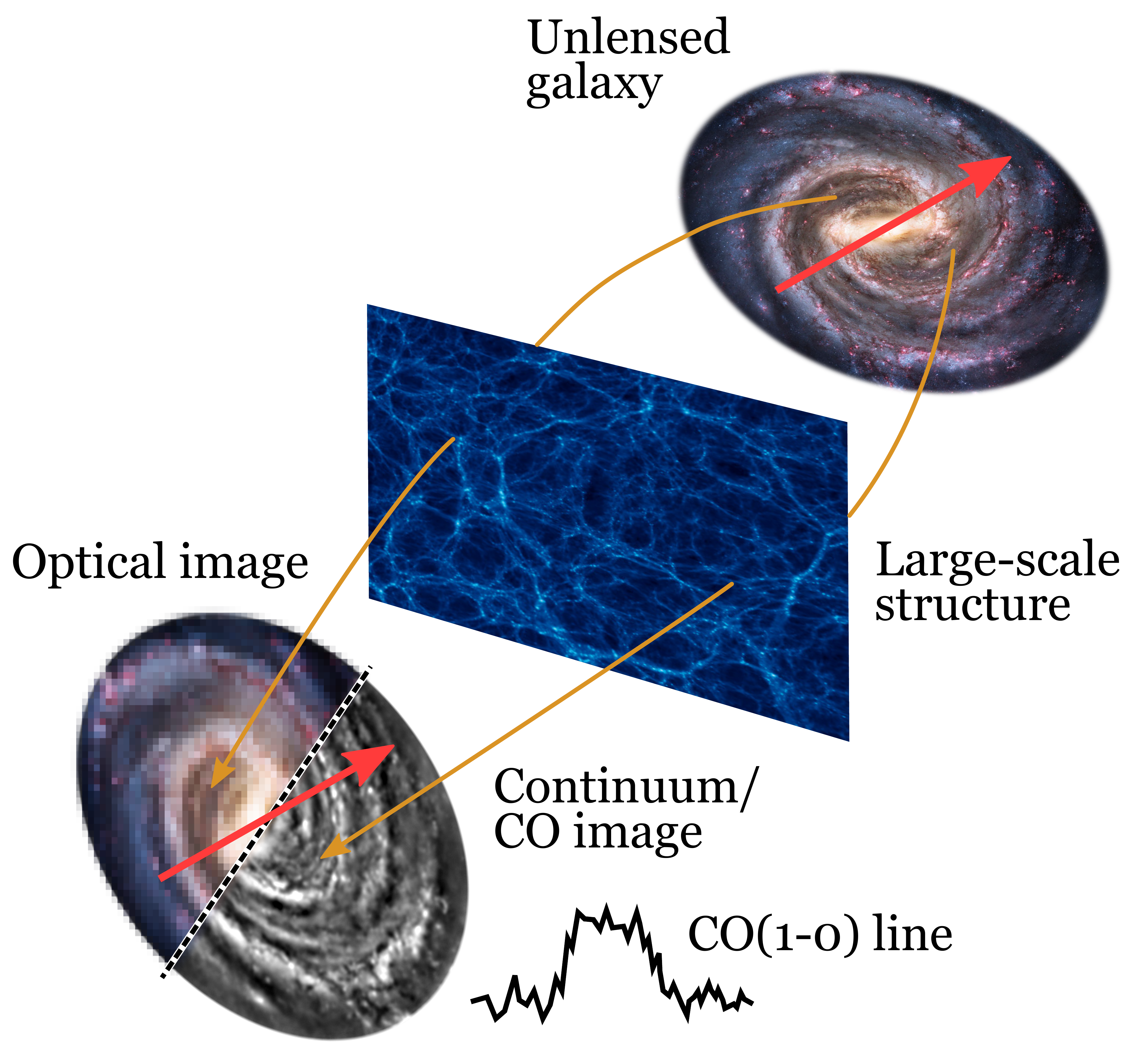}{fig:lensing}{Illustration showing the weak gravitational lensing of CO galaxies. Light rays (orange lines) are bent and sheared as they pass through the large-scale structure, resulting in a distortion of the observed shape of the galaxy. The galaxy can be imaged in multiple bands, including optical and radio continuum/CO line emission. The advantage of the CO/radio continuum strategy advocated here is that it preserves redshift information through detection of the CO(1--0) line, and has different PSF characteristics from the optical. The fact that lensing leaves the polarization angle unaffected (red arrows) can also be used to reduce shape noise and solve for the intrinsic alignment effect. (Composite image. Credit: P.~Bull/NASA/S.~Gottlober/MultiDark.)}

This information does not come easily however. Weak lensing observations remain some of the most challenging in all of astronomy, with precise control over myriad instrumental and astrophysical systematics required to even detect the effect, let alone measure its statistical properties with the sub-percent precision required to beat down errors on cosmological parameters. Despite a series of enhancements in analysis techniques \citep[e.g.][]{Kaiser:1994jb, Bernstein:2001nz, Miller:2007an, Huff:2017qxu}, low-level systematics can still remain in the data, subtly biasing the resulting cosmological constraints \citep{Heavens:2000ad, Hirata:2003cv, Huterer:2005ez, 2013MNRAS.429..661M, 2017arXiv171003235M}.

The development of a rigorous and independent cross-check of the optical weak lensing signal is therefore of vital importance -- without it, the unprecedented precision of flagship datasets such as the LSST lensing sample could be undermined. Radio weak lensing (Fig.~\ref{fig:lensing}) represents a highly promising solution to this problem \citep{Chang:2004ys, 2012arXiv1206.4437B, 2016arXiv160207482P, Camera:2016owj} -- the galaxy shape measurement process is fundamentally different in the radio, owing to the almost-deterministic nature of synthesized beams, the lack of atmospheric seeing effects, and differences in galaxy morphologies between the radio and optical \citep{2010MNRAS.401.2572P, Tunbridge:2016oio}. As such, many weak lensing systematics are expected to be strongly suppressed by radio-optical cross-correlations \citep{2016MNRAS.456.3100D}. A consistent picture between the radio and optical cross- and auto-power spectra would also be strong evidence that systematics are under control, and could therefore thoroughly validate the weak lensing methodology.

In addition to providing a cross-check on optical surveys, radio surveys also open up several exciting possibilities to measure extra information about the lensing galaxy sample that would be difficult, if not impossible, to obtain in the optical. Spectroscopic redshifts can be measured for large fractions of the sample for example, an extremely costly exercise in the optical. As well as removing photometric redshift calibration as a key source of uncertainty, this would allow more information to be recovered from the lensing signal in the radial direction, potentially significantly improving constraints on some cosmological parameters. Polarization information can also be readily obtained in the radio, allowing the intrinsic orientations of galaxies to be inferred -- again removing an important source of systematic error (intrinsic alignments).

In this chapter, we examine the possibility of measuring the weak lensing shear signal from a large sample of CO-emitting galaxies in several spectroscopic redshift bins across a substantial sky area and redshift range. This requires a survey of a large number of moderately-resolved CO-emitting galaxies. The galaxy shapes can be measured using continuum emission integrated over a few wide channels in each band, while the galaxy redshifts will be determined from spectroscopic detections of the CO(1-0) line. The resulting shear catalog can be cross-correlated with optical catalogs to greatly reduce the impact of systematics. We also discuss some of the novel analysis techniques that are enabled by radio observations, including polarization-based measurements of intrinsic alignments, and kinematic lensing using resolved spectral lines.



\section{Systematics mitigation with radio-optical cross-correlations} \label{sec:systematics}
Cosmological constraints from weak lensing surveys require the creation of a shear map on the sky. The spatially varying shear $\gamma$ must be estimated from the shapes of individual galaxies which sample the shear field, and is typically done using a simple (often weighted) average over galaxies in a sky patch small enough such that the shear is constant within the patch. Because weak lensing causes such a small change in the ellipticity of an individual galaxy, even small systematic biases in the estimation of its shape can quickly overwhelm the cosmological signal and cause significant errors in e.g. measuring the dark energy equation of state. For a linear model of bias on shear $\gamma_{\rm meas}  = (1 + m)\gamma_{\rm true} + c$ it is necessary for $m < \mathcal{O}(10^{-2})$ and $c < \mathcal{O}(10^{-3})$ for Stage III surveys such as DES and SKA1 and $m < \mathcal{O}(10^{-3})$ and $c < \mathcal{O}(10^{-4})$ for Stage IV surveys such as LSST and SKA2 \citep{Amara:2007as}.

Weak lensing with interferometers at radio wavelengths provides a unique opportunity to mitigate many of these systematics. Cosmological constraints are typically derived by taking two-point statistics of shear maps; if these power spectra are made by cross-correlating radio and optical shear maps, wavelength-dependent systematics either disappear or can be calibrated out \citep[see][]{Camera:2016owj}. Some of the most important systematics for optical surveys are as follows:

\paragraph{PSF systematics}
Estimating $\gamma$ from galaxy shapes requires deconvolution of the shape of the Point Spread Function (PSF) from each individual source. For ground-based optical surveys, atmospheric seeing produces a PSF which spatially varies in a non-deterministic way. Significant efforts are put into PSF reconstruction and strategies for precise and accurate deconvolution. Even for space-based surveys, challenges remain due to instabilities in the optical systems, non-trivial color dependence of the PSF, and effects within the CCD \citep{Mandelbaum:2015qia}.

In contrast, interferometer dirty beams may be seen as a stable and highly deterministic forward model, allowing for either precise determination of galaxy shapes in images or directly in the $uv$ plane. Even if shape measurement systematics remain, they will be highly uncorrelated with those from optical surveys, meaning systematic uncertainties from PSF removal should be highly suppressed in radio-optical cross-correlations.

\paragraph{Blended (confused) sources} A substantial fraction of the galaxies detected in deep optical surveys like LSST are expected to be overlapping on the sky. Such `blended' sources can be hard to identify, especially when the resolution limits imposed by atmospheric seeing are taken into account. Unidentified blends mostly add random noise to the measured shear signal, as the blended galaxies should have no preferred orientation. This source of noise can be removed simply by removing the blended sources (if they can be identified), although this can substantially reduce the effective number density of galaxies in the sample. More problematic is the effect of blends on photometric redshift estimation; trying to fit a photometric redshift to two sources at very different redshifts can result in catastrophic failure of the photo-z algorithm, which can ultimately cause biases in cosmological parameter estimates \citep{2018MNRAS.475.4524S}. The high angular resolution afforded by the long baselines of ngVLA may be useful in at least identifying blended sources for some of the galaxies in overlapping optical surveys from continuum images, while the degree of blending/confusion in the radio survey will only depend on instrumental resolution (since the atmospheric seeing is a much smaller effect). The availability of CO line redshifts will also help mitigate catastrophic photometric redshift errors caused by blends. The extent to which ngVLA data is useful for identifying blends in optical data will depend on the final array layout and depth of the survey, but if suitable optimizations are made, the impact could be considerable -- especially if photo-z estimation algorithms are unable to reduce their sensitivity to blends.

\paragraph{Intrinsic alignments}
The simple average estimator for $\gamma$ relies on the assumption that galaxy shapes were uncorrelated before lensing (i.e. in the absence of lensing, the average ellipticity should be zero). However, the realities of galaxy formation processes mean that this is not true in the real Universe. Failing to account for these intrinsic galaxy alignments may bias cosmological parameter estimation by many standard deviations \citep{Troxel:2014dba}. By measuring both the lensed total intensity galaxy position angle and a quantity which is not affected by lensing, such as polarization position angle \citep{2011MNRAS.410.2057B} or spatially-resolved kinematics \citep{Morales:2006fq}, intrinsic alignments can be mapped and accounted for in the lensing analysis. Being able to account for this directly has significant advantages over current strategies applied in optical weak lensing surveys, which typically involve either removing data (seriously degrading cosmological constraints) or the application of poorly-motivated models for the intrinsic alignment signal. Such a map of intrinsic alignments would not only be useful for an ngVLA weak lensing survey itself, but could also be used to model the intrinsic alignments within optical surveys (if the same galaxy populations are being probed).

\section{Limitations of current astronomical instrumentation}

While pathfinder weak lensing observations have been made with existing radio facilities \citep[e.g.][]{Chang:2004ys, 2010MNRAS.401.2572P}, they are far from achieving parity with optical surveys -- by any metric. The current state of the art is SuperCLASS\footnote{\url{http://www.e-merlin.ac.uk/legacy/projects/superclass.html}}, a $\sim 1.75$ deg$^2$ survey on e-MERLIN and JVLA that targets several Abell clusters with overlapping Subaru imaging. While these instruments have comparatively slow survey speeds, somewhat sparse $uv$ coverage (even when combined), and operate in a band ($\sim 1.4-1.5$ GHz) that lacks emission lines suitable for providing redshift information, SuperCLASS is nevertheless useful as a way of proving the radio weak lensing method and providing a testing ground for future techniques.

The Square Kilometre Array (SKA) and its precursors MeerKAT and ASKAP have better sensitivity and much denser $uv$ coverage, making imaging and shape measurement much easier. The precursors lack sufficiently long baselines that are needed to resolve higher-redshift continuum galaxies however, and so the first large-area radio weak lensing survey will probably have to wait until SKA1 comes online. A $\sim 5,000$ deg$^2$ continuum survey on SKA1-MID Band 2 (950 -- 1750 MHz) is expected to be competitive with the Dark Energy Survey \citep{Brown:2015ucq, 2016MNRAS.463.3674H}, and a fraction of the detected galaxies will have 21cm line detections and therefore redshifts. Unless they can be cross-matched with optical galaxies (with spectroscopic or photometric redshifts), the large fraction of sources without a 21cm detection will have effectively no redshift information, resulting in what would effectively be a 2D lensing survey (containing significantly less information about cosmological evolution).

As discussed in the previous section, optical surveys also have some inherent limitations when it comes to measuring the weak lensing signal, mostly centered around uncertainties in the shape, stability, and frequency-dependence of the point spread function, and a lack of knowledge about the intrinsic shapes of the galaxies. A number of these can be mitigated by some features of ngVLA and radio weak lensing observations more generally, as we discuss in the next section.

\section{Connection to unique ngVLA capabilities} \label{sec:unique}

The ngVLA will have a unique combination of spectral/spatial resolution and wide bandwidth, allowing it to detect large numbers of semi-resolved galaxies in both continuum and CO line emission at low and intermediate redshifts. Other radio weak lensing surveys have been proposed, for example using the SKA at lower frequencies ($\sim 1.5$ GHz), but these mostly lack redshift information except for a small fraction of sources that also have a detectable 21cm line. A CO galaxy sample would also be better matched to the LSST and DES samples than a 21cm sample, as it will preferentially contain normal star-forming galaxies that will make up the bulk of the optical samples. The higher angular resolution of ngVLA will also reduce the occurrence of blended/confused sources, which are a significant source of noise in lensing measurements \citep{2018MNRAS.475.4524S}.

ngVLA will also be able to operate in fundamentally different survey modes from optical experiments. A major advantage over optical telescopes is the ease with which polarized observations can be made. Weak lensing leaves the polarization angle of a galaxy unchanged. Since polarization traces the disk, this can be used to measure the intrinsic orientation of the galaxy, i.e. its orientation before it was sheared.
As discussed in Section~\ref{sec:systematics}, this helps to mitigate the effects of intrinsic alignments, which are correlations in the ellipticities of galaxies caused by non-lensing effects like tidal shears that are a significant nuisance effect for weak lensing cosmology \citep[e.g.][]{Joachimi:2015mma}.

The high spectral resolution of ngVLA can be used to similar effect. If the CO lines of some subset of galaxies can be spectrally resolved, one can obtain highly complementary information about their circular velocity. As in the polarized case, this can be used to infer the intrinsic ellipticity and thus greatly suppress shape noise \citep{Morales:2006fq}. If a significant decrease in shape noise can be achieved, the number density of galaxies required to obtain a high SNR detection of the lensing signal can also be greatly reduced, making ngVLA a much more efficient weak lensing survey instrument without the need to substantially alter its design.

\articlefigure[width=0.7\textwidth]{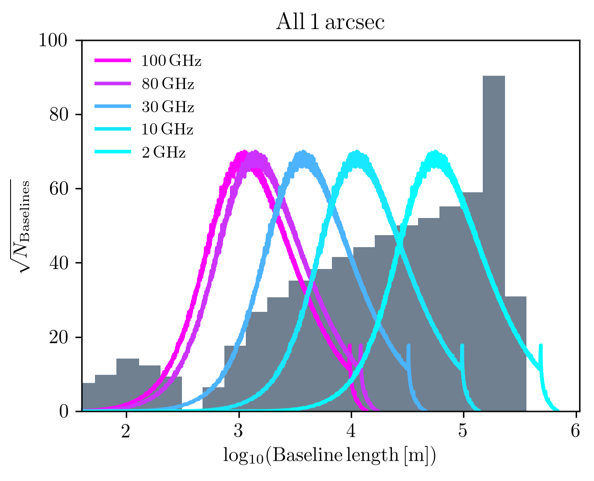}{fig:shear}{Comparison of the angular resolution of ngVLA to the angular scales affected by the cosmic shear signal, as a function of frequency band (redshift). The colored curves show the (normalized) shear signal for different observed CO(1--0) line frequencies. They were calculated by taking a representative field of view populated with galaxies, smoothed to 1 arcsec resolution, applying an appropriate level of weak lensing shear, and then calculating the power spectrum of the difference between the sheared and unsheared images (note the Nyquist artefact in the tails). The gray histogram is the orientation-averaged ngVLA baseline distribution, assuming the Conway array layout.}

\section{Measurement technique and experimental layout}

In this section, we give a high-level overview of how a CO weak lensing survey could be carried out in practice. Two measurements are required per galaxy:
\renewcommand{\theenumi}{(\alph{enumi})}
\begin{enumerate}
    \item The galaxy's shape (ellipticity), which can be obtained with high signal-to-noise through imaging of the continuum emission;
    \item A spectroscopic determination of its redshift through a detection of the CO (1-0; 115.2712 GHz) line.
\end{enumerate}
The first measurement requires that the target galaxies be sufficiently resolved that their ellipticities can be measured. This suggests a target resolution of $\sim$1'' for galaxies at $z \sim 1$ (see Fig.~\ref{fig:shear}). A wide instantaneous bandwidth is preferred to maximize continuum SNR, ideally breaking up the full available bandwidth into a handful of wide channels to allow for the correction of frequency-dependent beam effects. Note that the galaxies need only be semi-resolved in order to measure their ellipticities accurately enough. Beam smearing effects will be important when measuring the shape of the galaxy, hence the need to break up the band into smaller chunks; the PSF must be known very accurately, including its frequency dependence, and time-variation of the PSF should also be strongly controlled. An accurate flux density calibration is not strictly required, except to apply cuts to select certain galaxy sub-populations. Imaging may not be strictly necessary if visibility-space shape measurement methods can be used \citep{2002ApJ...570..447C, 2014MNRAS.444.2893P, 2016MNRAS.463.1881R}, although this will require gridded visibilities to be stored \citep{2015arXiv150706639H}.

The second measurement requires just enough spectral resolution to identify and measure the redshifted CO line integrated over the solid angle of the galaxy. In the simplest case, there is no requirement to accurately measure the line flux, line width, or other spectroscopic properties, so a low-SNR line detection is sufficient. Depending on typical linewidths, the spectral resolution required will likely be a few hundred km/s.

The SNR of the weak lensing measurement strongly depends on the number density of sources for which shapes can be measured, as well as the spatial volume that can be surveyed. A minimal proof of concept survey would need to target a few hundred square degrees with source number densities of order a few per sq. arcmin. An ideal survey for studying lensing systematics would be matched to the WFIRST footprint ($\sim$2,200 sq. deg.), with a high source number density ($\sim 60-70$ per sq. arcmin). An ideal cosmological survey would cover a large fraction of the Euclid or LSST footprints ($\sim$15,000 sq. deg.) with 20-30 sources per sq. arcmin. Lower number densities could be sustained with the addition of spectroscopic redshift information, which compensates by providing access to radial Fourier modes that LSST and other photometric surveys effectively discard. A continuum-only weak lensing survey would also be possible, negating the need to carry out measurement (b), but is unlikely to be competitive with optical surveys with photometric redshifts.



An alternative to increasing the source number density is to measure (c) the polarization angle of the galaxies; or (d) accurately measure their linewidths. These would allow the methods described in Section~\ref{sec:unique} to be used to recover the intrinsic galaxy shapes/alignment, thus significantly cancelling important noise sources such as shape noise and intrinsic alignments. Carrying out these much more detailed measurements would likely necessitate a much smaller survey area, with greater depth per pointing. The trade-offs between these different survey modes will need to be explored within the context of a realistic model of the (polarized) luminosity, size, and circular velocity distribution of CO galaxies.

Finally, a significant hurdle is the small field of view of ngVLA, which substantially reduces its survey speed. The $\sim 1$ deg primary beam of SKA gives it a significant advantage in this respect. Array designs that increase the field of view of ngVLA and optimize its baseline distribution for sensitivity at $\sim 1$ arcsec scales (c.f. Fig.~\ref{fig:shear}) are therefore highly desirable for this science case. Examples of how to achieve this might include reducing dish sizes and adopting a more compact layout with a denser core, perhaps in the context of a `hybrid' design with a dense core of small dishes at the center and larger dishes spread out to longer baseline lengths. An advantage of such a design is that the core and outer sets of dishes could be operated independently if desired, allowing large surveys and high-resolution pointed observations to be performed concurrently.

\section{Complementarity}

While a radio weak lensing survey performed with ngVLA on its own would be a interesting and potentially valuable cosmological test, the power of performing such a survey lies firmly in the ability to jointly analyze it with an overlapping optical survey. The cross-correlation of a radio weak lensing sample with contemporary optical datasets (e.g. from LSST, WFIRST, and Euclid) would provide an extremely robust validation of the weak lensing methodology that so much of the observational cosmology program will rely upon in the next decade. This represents an immensely valuable synergy with practically all of the large cosmological survey instruments planned for the beyond-2025 timeframe, and would represent the ``last word'' in the interpretation of weak lensing data.

This science case also complements and extends the radio weak lensing science case of the SKA to higher frequencies, higher spatial resolutions, and arguably a more appropriate sample for optical cross-correlations, as well as providing high-value spectroscopic redshift information, and the ability to use novel techniques such as kinematic lensing and intrinsic alignment removal with polarization. Additionally, the large sample of galaxies that would be obtained in performing a radio weak lensing survey would clearly be valuable for a wide array of astrophysical applications too, such as studies of galaxy formation and evolution, and galaxy structure and dynamics (if spectrally resolved/polarized observations are prioritized), providing a further connection to future NASA flagship missions such as the Origins Space Telescope and LUVOIR.

\acknowledgements We are grateful to J. Lazio for useful discussions.



\end{document}